Research Article

# Detailed Derivation of the Scalar Explicit Expressions Governing the Electric Field, Current Density, and Volumetric Power Density in the Four Types of Linear Divergent MHD Channels Under a Unidirectional Applied Magnetic Field

Osama A. Marzouk

College of Engineering, University of Buraimi, Al Buraimi, Postal Code 512, Oman
E-mail: osama.m@uob.edu.om

**Abstract:** The current study belongs to the field of applied mathematics in plasma physics and electric power, where mathematical analysis of the algebraic equations governing the electric field vector, and the electric-current density field vector within a Magnetohydrodynamic (MHD) linear two-dimensional divergent supersonic channel is utilized to derive analytical expressions for these important fields, as well as closed-form equations for the volumetric power density (output electric power per unit volume of the plasma channel). The expressions presented here describe analytically the operation of the MHD channel as an electric power source within an Open-Cycle Magnetohydrodynamic (OCMHD) generator. The four common types of the MHD linear channels are covered here: namely, (1) continuous-electrode Faraday channel, (2) linear Hall channel, (3) segmented-electrode Faraday channel, and (4) diagonal-electrode channel. The mathematical results, their detailed derivation, and the companion graphical illustrations aid in making a proper decision regarding which channel type is the most suitable for a given application. Under typical operational conditions of 5 S/m plasma electric conductivity, 5 T magnetic field, and 2,000 m/s plasma speed, as well as an optimized load factor of 0.5, we estimate the following numerical values (unsigned magnitudes) for the continuous-electrode Faraday channel (with a Hall parameter of 1): useful electric field (across the external electric load): 5 kV/m, useful electric current-density (between the terminal electrodes within the channel): 12.5 kA/m$^2$, volumetric power density (dissipated by the load per unit volume of plasma): 62.5 MW/m$^3$, and electric efficiency (for the electric field or voltage): 50%. For the Hall linear channel (with a Hall parameter of 5), these quantitative performance values become 25 kV/m, 4.808 kA/m$^2$, 120.19 MW/m$^3$, and 46.30%. For either an infinitely segmented-electrode Faraday channel or an on-design diagonal-electrode linear channel (and irrespective of the Hall parameter), these quantitative performance values become 5 kV/m, 25 kA/m$^2$, 125 MW/m$^3$, and 50%.

# Abbreviation

| | |
|---|---|
| $\beta$ | Hall parameter [dimensionless] |
| $\delta$ | Distance for approximating the gradient [m] |
| $\Phi$ | Electric potential; voltage [V] |
| $\gamma$ | Specific heat ratio [dimensionless] |
| $\eta_{\text{elec}}$ | Electric efficiency [%] |
| $\eta_{\text{elec}, D}$ | Electric efficiency in a diagonal-electrode channel with uniform conditions, when operating at its design point [%] |
| $\eta_{\text{elec}, F-\text{cont}}$ | Electric efficiency in a continuous-electrode Faraday channel with uniform conditions [%] |
| $\eta_{\text{elec}, F-\text{seg}}$ | Electric efficiency in a segmented-electrode Faraday channel with uniform conditions [%] |
| $\eta_{\text{elec}, H}$ | Electric efficiency in a linear Hall channel with uniform conditions [%] |
| $\eta_{\text{isen}}$ | Isentropic efficiency [%] |
| $\mu$ | Mobility of electrons [(m/s)/(V/m) or m$^2$/(V.s)] |
| $\sigma$ | Electric conductivity [S/m or C/(m.V.s)] |
| $\theta_\Phi$ | Tilt angle (acute angle, measured from the $y$-axis) of the equipotential lines [°, degree] |
| $\theta_J$ | Tilt angle (acute angle, measured from the $y$-axis) of the current-density vector [°, degree] |
| $A$ | Area [m$^2$] |
| $B$ | Unidirectional applied magnetic-field flux density (external magnetic field) [T, tesla] |
| $\vec{B}$ | The applied magnetic field when expressed as a vector (pointing in the positive $z$-axis) [T, tesla] |
| $dV$ | Volume element (for three-dimensional integration) [m$^3$] |
| $\vec{E}$ | The electric field within the plasma channel (as felt by the plasma gas) [V/m] |
| $\vec{E}_0$ | The electric field consumed by the external load [V/m] |
| $\vec{E}_{OC}$ | The induced open-circuit electric field at the channel electrodes (when no external load is connected) [V/m] |
| $E_x$ | The axial ($x$-component) of the internal (within the channel) electric field vector $\vec{E}$ [V/m] |
| $E_{0x}$ | The axial ($x$-component) of the electric field outside the channel (as consumed by the external electric load) [V/m] |
| $E_y$ | The vertical ($y$-component) of the electric field vector $\vec{E}$ [V/m] |
| $E_{0y}$ | The vertical ($y$-component) of the electric field outside the channel (as consumed by the external electric load) [V/m] |
| $E_{OC}$ | Open circuit (if no external load) induced electric field when approximated as a scalar (unidirectional) quantity. For Faraday-type channels (continuous-electrode Faraday channel or segmented-electrode Faraday channel), $E_{OC} = u\,B$. For Hall-type channels, $E_{OC} = \beta\,u\,B$. |
| $EE$ | Enthalpy extraction [%] |
| $e$ | Elementary charge [C] |
| $h$ | Specific static enthalpy [J/kg] |
| $h_0$ | Specific total (stagnation) enthalpy [J/kg] |
| $I$ | Conventional electric current (whose direction is determined assuming positive charge carriers) that flows externally outside the Magnetohydrodynamic (MHD) generator to the electric load [A] |
| $\hat{i}$ | Unit vector along the positive $x$-axis [dimensionless] |
| $\vec{J}$ | Conventional electric current-density vector (whose direction is determined assuming positive charge carriers) within the plasma channel [A/m$^2$] |
| $J_x$ | Axial ($x$-component) of the conventional electric current-density vector $\vec{J}$ [A/m$^2$] |
| $J_y$ | Vertical ($y$-component) of the conventional electric current-density vector $\vec{J}$ [A/m$^2$] |
| $\hat{j}$ | Unit vector along the positive $y$-axis [dimensionless] |
| $K_F$ | Load factor for Faraday-type channels (continuous-electrode Faraday channel or segmented-electrode Faraday channel) [dimensionless]. This load factor ranges from 0 (short-circuited load) to 1 (open circuit, disconnected load). For either extreme case, the power dissipated by the load is zero (because of having zero voltage drop in the short circuit case, or having zero current flow in the open circuit case). We have $K_F = \dfrac{R_L}{(R_L + R_G)}$. |
| $K_H$ | Load factor for Hall-type channels. It ranges from 0 (short-circuited load) to 1 (open circuit, disconnected load) [dimensionless]. For either extreme case, the power dissipated by the load is zero (because of having zero voltage drop in the short circuit case, or having zero current flow in the open circuit case). We have $K_H = \dfrac{R_L}{(R_L + R_G)}$. |
| $\hat{k}$ | Unit vector along the positive $z$-axis [dimensionless] |
| $n_e$ | Number density of electrons [m$^{-3}$] |
| $\hat{n}_B$ | Unit vector along the applied magnetic-field flux density vector [dimensionless] |
| $P$ | Power density or volumetric power density (output electric power per unit volume of plasma) [W/m$^3$] |
| $P_D$ | Output electric power per unit volume of plasma in diagonal-electrode Faraday channels [W/m$^3$] |
| $P_{D,\,\text{opt}}$ | Output electric power per unit volume of plasma in diagonal-electrode channels with a matched (optimized) load [W/m$^3$] |
| $P_F$ | Output electric power per unit volume of plasma in Faraday channels [W/m$^3$] |
| $P_{F-\text{cont}}$ | Output electric power per unit volume of plasma in continuous-electrode Faraday channels [W/m$^3$] |
| $P_{F-\text{cont, opt}}$ | Output electric power per unit volume of plasma in continuous-electrode Faraday channels with matched (optimized) load [W/m$^3$] |
| $P_{F-\text{cont, opt, ideal}}$ | Output electric power per unit volume of plasma in continuous-electrode Faraday channels with a matched (optimized) load at the limiting case of $(\beta \to 0)$ [W/m$^3$] |
| $P_{F-\text{seg}}$ | Output electric power per unit volume of plasma in segmented-electrode Faraday channels [W/m$^3$] |
| $P_{F-\text{seg, opt}}$ | Output electric power per unit volume of plasma in segmented-electrode Faraday channels with a matched (optimized) load [W/m$^3$] |
| $P_H$ | Output electric power per unit volume of plasma in linear Hall channels [W/m$^3$] |
| $P_{H,\,\text{opt}}$ | Output electric power per unit volume of plasma in linear Hall channels with a matched (optimized) load [W/m$^3$] |
| $P_{\text{opt}}$ | Output electric power per unit volume of plasma with a matched (optimized) load [W/m$^3$] |
| $P_w$ | Output electric power from a magnetohydrodynamic channel [W] |
| $p_0$ | Total (stagnation) pressure [Pa, N/m$^2$] |
| $R$ | Resistance of a solid conductor (in general) [$\Omega$ or S$^{-1}$] |
| $R_L$ | External resistance (of the load) [$\Omega$ or S$^{-1}$] |
| $R_G$ | Internal resistance (of the plasma generator) [$\Omega$ or S$^{-1}$] |
| $u$ | Unidirectional axial plasma bulk speed [m/s] |
| $\vec{u}$ | The unidirectional axial plasma bulk speed when expressed as a vector (pointing in the positive $x$-axis) [m/s] |

# 1. Introduction
## 1.1 *Background*

Electricity is a main driver for modern civilization [1, 2]. Global electricity generation and demand have been generally increasing steadily since 2000, except for irregularities in 2009 due to the 2008 economic recession and in 2020 due to the COVID-19 pandemic [3–5]. While conventional power systems (i.e., power plants that burn fossil fuels [6–8]) are reliable methods for generating large amounts of electricity, their harmful Greenhouse Gas (GHG) emissions [9–12] and other pollutants encourage the shift to non-conventional alternatives, such as renewable energy sources (particularly solar energy [13–16] and wind energy [17–19]), and low-carbon hydrogen [20, 21].

Another method for non-conventional electricity generation is the Magnetohydrodynamic (MHD) generator. Such generators utilize the Lorentz force and electromagnetic principles to extract energy from a moving ionized gas (plasma) to produce Direct Current (DC) electricity without rotating or reciprocating elements as found in conventional heat engines [22–24]. The electrically conductive medium in MHD generators can be an ionized gas mixture (plasma) or a liquid metal [25–27]. We here limit ourselves to gas-based MHD generators because we focus on plasma formed from gaseous combustion products [28–30], which is a popular and mature process in fuel-fired thermal power plants.

There are two broad categories of MHD generators. One category of MHD generators is the open-cycle generators, where combustion products (such as carbon dioxide "$CO_2$" and steam "$H_2O$") are seeded with alkaline compounds (e.g., potassium carbonate "$K_2CO_3$"), and they form weakly-ionized thermal (equilibrium) plasma that is accelerated within a channel while subject to a magnetic field to induce electric fields and electric currents [31, 32]. Such a channel may be viewed as an "electromagnetic turbine". Being in thermal equilibrium (or thermodynamic equilibrium) means that the electrons (the effective charge carriers due to their light mass and thus high mobility) and the heavy particles in the plasma gas (ions, atoms, and molecules) in such combustion plasmas can be treated as having one common temperature due to the high collision frequencies and high energy transfer per collision [33, 34]. Being open-cycle means that the working plasma is not recirculated back after energy is extracted from it for electricity generation.

The other broad category of MHD generators is the closed-cycle type, in which MHD generators utilize a heated inert gas (e.g., argon "Ar") [35, 36] whose temperature is elevated using a heat exchanger (rather than a direct combustion process), and this heated gas is seeded with an alkaline metal, such as cesium vapor (Cs) to form a non-equilibrium plasma medium. It should be noted here that the alkali metal cesium, which has an atomic number of 55, vaporizes at a relatively low temperature of about 944 K (671 °C) [37, 38], while it readily melts near room temperatures at only 302 K (29 °C) [39, 40]. Being a non-equilibrium plasma means that its constituent gas particles (electrons, ions, atoms, and molecules) are not in a state of thermal equilibrium, with electrons characterized by a higher temperature compared to heavy particles (ions, atoms, and molecules) [41, 42]. The higher temperature of electrons in inert gas plasmas is necessary for keeping a sufficient level of electric conductivity despite the relatively low bulk temperatures. This state of non-equilibrium (two-temperature plasma) can be achieved by ohmic heating (Joule heating) under a relatively small frequency of collisions between the electrons and the neutral gas atoms [43, 44]. Being closed-cycle means that the working plasma is recirculated back after energy is extracted from it for electricity generation. Before repeating the energy extraction process, energy is added again to the depleted plasma (e.g., by a heat exchanger) to restore its initial state [45].

Closed-Cycle Magnetohydrodynamic (CCMHD) generators typically have a disc shape [46–50], and this design allows efficient utilization of the magnets [51]. The compactness of CCMHD generators renders them of special interest in space applications, especially in space missions reaching trajectories far from the sun, where solar irradiation and photovoltaic power conversion become ineffective [52].

On the other hand, Open-Cycle Magnetohydrodynamic (OCMHD) generators pertain more to terrestrial electricity generation, where they can provide much larger production capacities similar to utility-scale power plants that use an open (non-recirculating) Brayton cycle or a closed (circulating) Rankine cycle [53–55]. Such OCMHD generators may be used either in a continuous mode as power plants or in a pulsed mode for geological prospecting (like the Sakhalin 510 MW unit [56, 57]).

In a recent review study of MHD power generation, one of the mentioned advantages of OCMHD was the ability to work at very high temperatures (such as 2,800 K) that are well beyond what conventional turbines of any type can tolerate. From the Carnot efficiency condition in thermodynamics [58, 59], such higher temperatures theoretically mean that higher thermal efficiencies of energy conversion are possible, provided that the plasma can be exploited to near ambient temperatures. That review study also referred to a number of technical challenges for MHD power generation, which included improving the scalability of that MHD technology [60].

The performance of MHD generators can be described in terms of three percentage metrics [61, 62].

First, we have the enthalpy extraction or Enthalpy Extraction ratio ($EE$) metric, which is the ratio of the drop in static enthalpy ($\Delta h$) within the MHD generator to the inlet static enthalpy ($h_{in}$) at the entrance of the MHD generator. Thus,

$$EE = \frac{\Delta h}{h_{\text{in}}} = 1 - \frac{h_{\text{out}}}{h_{\text{in}}} \tag{1}$$

where ($h_{\text{in}}$) and ($h_{\text{out}}$) are the specific static enthalpy (which is a thermodynamic property that expresses energy content per unit mass of a fluid medium [63]) of the plasma at the entrance and at the exit of the MHD generator. A high Enthalpy Extraction (*EE*) of 38% was successfully demonstrated experimentally for a disc-type MHD generator [64].

Second, there is the MHD generator isentropic efficiency ($\eta_{\text{isen}}$) [65], which relates the actual relative drop in the static enthalpy to the ideal relative drop in the total (stagnation) enthalpy ($h_0$) in the case of an isentropic process under the assumption of a calorically-perfect gas (thus, having constant specific heat capacities and constant specific heat ratio) [66–68]. This MHD isentropic efficiency is

$$\eta_{\text{isen}} = \frac{\frac{\Delta h}{h_{\text{in}}}}{\frac{\Delta h_0}{h_{0,\,\text{in}}}} \tag{2}$$

where ($h_{0,\,\text{in}}$) is the specific total or stagnation (which refers to the hypothetical condition when the fluid becomes at rest [69, 70]) enthalpy of the plasma at the entrance of the MHD generator.

The above equation can be written in terms of the total pressure (the stagnation pressure, which is the sum of the static pressure and the dynamic pressure [71, 72]) at the entrance of the MHD generator ($p_{0,\,\text{in}}$) and at the exit of the MHD generator ($p_{0,\,\text{out}}$) using standard expressions for isentropic compressible calorically-perfect flows [73] as

$$\eta_{\text{isen}} = \frac{\frac{\Delta h}{h_{\text{in}}}}{1 - \left(\frac{p_{0,\,\text{out}}}{p_{0,\,\text{in}}}\right)^{\frac{\gamma-1}{\gamma}}} \tag{3}$$

where ($\gamma$) is a constant specific heat ratio (ratio of the specific heat capacity at constant pressure to the specific heat capacity at constant volume [74, 75], and it is also called the adiabatic index [76]). A high MHD isentropic efficiency ($\eta_{\text{isen}}$) of 63% was successfully demonstrated experimentally for a disc-type MHD generator [77].

Third, there is the MHD generator electric ($\eta_{\text{elec}}$), which is defined as the ratio of the electric power to the mechanical power. Mathematically, this is expressed as

$$\eta_{\text{elec}} = \frac{-\iiint_{\text{volume}} \vec{J} \cdot \vec{E}_0 dV}{-\iiint_{\text{volume}} \vec{u} \cdot (\vec{J} \times \vec{B}) dV} \tag{4}$$

Open-Cycle Magnetohydrodynamic (MHD) generators are primarily in the form of a linear channel, in which the weakly-ionized plasma moves at a high speed while subject to a strong applied magnetic field [78]. The inlet plasma speed can be subsonic (Mach < 1) [79, 80] or supersonic (Mach > 1) [81–84]. Contrary to subsonic and incompressible flows [85, 86], supersonic flows at the MHD channel entrance need a divergent MHD channel in order to allow the supersonic flow to expand and accelerate [87–89]. Attached electrodes are used to transmit the collected electric current to an external eclectic load. There are four common designs for the open-cycle MHD channels; namely, (1) continuous-electrode Faraday channel [90–92], (2) segmented-electrode Faraday channel [93–95], (3) linear Hall channel (always have segmented electrodes) [96–98], and (4) diagonal channel or diagonal-electrode channel or diagonally-shorted channel (always have segmented electrodes) [99–101]. Each of these designs has a unique layout of its electrodes, hence the name of its type.

### 1.2 *Goal of the study*

This work can be viewed as an extension of our earlier studies about mathematical modeling for Magnetohydrodynamic (MHD) power generation. We previously presented the governing electric equations in the MHD channel under various levels of modeling approximation [102]. We also presented a proposed mathematical model for relating the electric conductivity of MHD plasma to its local temperature, pressure, and chemical composition [103].

In continuation of our work about mathematical modeling for Magnetohydrodynamic (MHD) generators, we here aim to present mathematical expressions for the electric field and electric current-density field within the plasma of an open-cycle MHD generator, and the resulting output power density under the four types of linear MHD channels. We supplement the mathematical analysis with illustrative sketches and logical derivation starting from elementary equations.

### 1.3 *Benefits of the study*

Our review of the mathematical expressions for linear Magnetohydrodynamic (MHD) generator channels, our organized presentation, and contrasting have a number of benefits. First, it serves as an introductory guide for readers interested in Open-Cycle MHD (OCMHD) power generation and the differences in their performance according to the different channel configurations. Second, our review provided here involves explicit algebraic scalar expressions, enabling parametric and visual investigation of the influence of some parameters on various variables of importance in the OCMHD channel. For example, the penalty in the electric output power density ($P$) due to the Hall parameter ($\beta$) in the case of a continuous-electrode Faraday channel can be easily identified through the presented mathematical expressions in the current study. Third, the directions of the vector fields under each channel configuration are clarified, and this is particularly useful in understanding the operation of the MHD generator for each case by a reader who is not an expert in plasma physics and MHD power generation. Fourth, the analytical treatment presented here facilitates optimizing the performance. For example, the load factor in the linear Hall channel can be optimized for maximum power output through the presented mathematical expressions in this study. Fifth, this study can be used in its exact form or can be modified and integrated with teaching materials or projects for enriching undergraduate academic courses in plasma physics, plasma applications, energy systems, and electrical power generation; and this can improve the conceptual understanding of students [104].

## 2. Research method

The method followed in the current study is the symbolic mathematical manipulation of a fundamental generic equation for electrically conducting moving plasma. This equation is the generalized (extended) Ohm's law in its vector form, which accounts for the Hall effect (Hall current-density) as well as the electromagnetic induction effect (Faraday current-density).

We expand this vector equation into three scalar equations (corresponding to the three Cartesian components). Then, we apply our assumptions to deduce two scalar equations for the axial component (along the plasma's travel direction) and the vertical components (perpendicular to the magnetic field and the plasma velocity). We drop the third scalar equation (in the direction of the magnetic field) because it becomes trivial in our case.

These two derived scalar equations are analyzed further for each of the four linear channel configurations; and this leads to the aimed mathematical expressions for the electric field, electric current-density, and power density for each configuration.

The common assumptions made in this study are:

The MHD channel has a divergent geometry, with a trapezoidal cross-section (in the $x-y$ plane). Such a simple MHD channel has been realized in the Sakhalin pulsed MHD generator [105]. The width (along the magnetic field) is constant, and its influence is disregarded here (this is equivalent to assuming infinite width, thus two-dimensional channels).

- The charge carriers are only the free electrons in the plasma (liberated as a result of thermal ionization). This means that while ions also exist (to ensure the overall neutrality of the plasma), their contribution to the electric current is neglected [106, 107]. This is a reasonable assumption given the much stronger mobility of the lighter electrons compared to the heavier ions [108–110].

- Unidirectional magnetic field (magnetic-field flux density) that points in the positive $z$-axis. Therefore, the magnetic-field flux density vector ($\vec{B}$) can be expressed as $B\,\hat{k}$, where $\hat{k}$ is a unit vector in the direction of the positive $z$-axis. Because the magnetic field is externally applied, this assumption can be justified. In such a case, special electromagnetic designs can be made to approximate this assumption. This treatment of the magnetic field as being fully controllable implies a low magnetic Reynolds number assumption [111–114], where auxiliary induced magnetic-field flux density due to the moving plasma (the self-excitation phenomenon) is neglected [115–118]. This "inductionless" assumption [119–121] of a low magnetic Reynolds number is reasonable for MHD generators [122–124].

- Unidirectional plasma velocity that points in the positive $x$-axis. Therefore, the plasma velocity vector ($\vec{u}$) can be expressed as $u\,\hat{i}$, where $\hat{i}$ is a unit vector in the direction of the positive $x$-axis. Although this assumption neglects turbulence and no-slip effects in the plasma flow, it can be regarded as an acceptable treatment for deriving system-level laws, where the time-averaged bulk velocity of the plasma should be primarily in the axial direction. This assumption becomes more valid when the divergence angle of the channel decreases, so the channel height approaches uniformity. In addition, turbulence tends to be suppressed as the Mach number increases [125, 126], and our study is for supersonic channels. In addition, adopting a one-dimensional approximation for a channel flow or exterior flow has been implemented in other studies [127–130].

- No electric field along the lateral direction (along the direction of the magnetic field). This assumption is aligned with the unidirectionality assumption for the magnetic field. Even if the plasma has a three-dimensional flow velocity, the unidirectional magnetic field along the $z$-axis is not able to induce an electric field in the same $z$-direction. Therefore, the electric field along the $z$-axis within the MHD plasma can only be caused by an externally applied electric field, but such a case is not considered in the current study, where there are no electrodes along the $z$-axis to permit this.

## 3. Results

In this section, we present a detailed derivation of scalar mathematical equations that describe various performance aspects of four types of linear MHD channels. We start with a common preparatory subsection, and then dedicate one subsection to each of the four channels. Therefore, the current section is divided into five subsections.

### 3.1 *Base scalar equations for electric fields in MHD plasma*

In this subsection, we derive the two base scalar equations for the conventional electric current-density and the load-consumed electric field, under the assumptions made in the previous section, particularly the unidirectional magnetic field and the unidirectional plasma velocity. These scalar equations are the $x$-component and the $y$-component of the generalized Ohm's law, which applies to a three-dimensional conductor medium (rather than a conductive wire).

The generalized Ohm's law (for plasma with neglected ion drift) relates the electric current-density ($\vec{J}$) to the electric field ($\vec{E}$), electric conductivity ($\sigma$), and the applied magnetic field ($\vec{B}$) [131–133]. It should be noted that we adopt the conventional current here, rather than the electron current [134–136]. The only difference if the electron current is adopted rather than the conventional current is that the electric current-density ($\vec{J}$) should have a minus sign in each appearance for it in our presented mathematical formulations.

The generalized Ohm's law can be expressed in different forms. We start with the following convenient form for it, which facilitates our discussion and subsequent analysis:

$$\vec{J} = \sigma \vec{E} - \mu \vec{J} \times \vec{B} \qquad (5)$$

where ($\mu$) is the electron mobility.

The Hall parameter ($\beta$) is the electron mobility ($\mu$) divided by the magnitude of the applied magnetic-field flux density ($B$). Mathematically, this is expressed as

$$\mu = \frac{\beta}{B} \qquad (6)$$

It is useful to add here that the electric conductivity of the plasma ($\sigma$) due to the free electrons (ions contribution is neglected as mentioned earlier) is [137–139]

$$\sigma = e\, n_e\, \mu \qquad (7)$$

where ($n_e$) is the number density of the charge-carrier electrons, and ($e$) is the elementary charge (the absolute electric charge of an electron; $e = 1.6021766 \times 10^{-19}\ C$ [140–142]).

Therefore, the generalized Ohm's law form in Equation (5) can be expressed in an alternative form as

$$\vec{J} = \sigma \vec{E} - \beta \vec{J} \times \hat{n}_B \qquad (8)$$

where ($\hat{n}_B$) is a unit vector in the direction of the applied magnetic-field flux density vector ($\vec{B}$). Thus, this unit vector is defined as

$$\hat{n}_B \equiv \frac{\vec{B}}{B} \qquad (9)$$

The first term in Equation (8), the term ($\sigma \vec{E}$), is an extended version of the classical Ohm's law for a one-dimensional solid conductor. To demonstrate this; one can multiply both sides of the equation by the area ($A$) perpendicular to the electric current-density, and then replace the vector electric field ($\vec{E}$) by the negative value of the scalar gradient of the electric potential ($\Phi$). This gives

$$J\,A = \sigma\,A \left( -\frac{\Delta \Phi}{\delta} \right) \qquad (10)$$

where ($J$) is the magnitude of the unidirectional current density; ($\delta$) is the separation distance between the two electrodes; and $\left( -\frac{\Delta \Phi}{\delta} \right)$ is a numerical approximation for the gradient of the electric potential ($\Phi$), and it becomes exact in the case of a linear decrease of the electric potential in the direction of the conventional electric current density.

The product ($J\,A$) in the above equation is the electric current ($I$).

$$I = J A \tag{11}$$

Also, the quantity $\left(\dfrac{\sigma A}{\delta}\right)$ is the reciprocal of the resistance ($R$) of the conductor. Therefore,

$$R = \dfrac{\delta}{\sigma A} \tag{12}$$

Using Equation (11) and Equation (12) in Equation (10) gives the classical Ohm's law, as

$$I = -\dfrac{\Delta \Phi}{R} \tag{13}$$

The minus sign in the above equation indicates that the electric potential drops in the direction of the electric current. This minus sign is normally dropped when Ohm's law is expressed, with the understanding that the electric potential drops across the conductor or the load.

The second term in Equation (8), the term $(-\beta \, \vec{J} \times \hat{n}_B)$, is the Hall electric current-density due to the Hall effect, where an electric voltage is induced as a result of the movement of the electrons under the effect of the applied magnetic field, and this leads to a "secondary" Hall current density perpendicular to both the "primary" current density ($\vec{J}$) and the applied magnetic field ($\vec{B}$) [143–145]. The magnitude of this "secondary" or Hall current density, $|-\beta \, \vec{J} \times \hat{n}_B|$, is proportional to the Hall parameter ($\beta$). Thus, from Equation (6), The magnitude of this "secondary" or Hall current density, $\left|-\beta \, \vec{J} \times \hat{n}_B\right| = |-\mu \, B \, \vec{J} \times \hat{n}_B|$, is proportional to either the magnitude of the applied magnetic field ($B$) or the electron mobility ($\mu$).

In the current study, the applied magnetic-field flux density vector is assumed to be totally in the direction of the $z$-axis, whose unit vector is ($\hat{k}$). Therefore, we have

$$\hat{n}_B = \hat{k} \tag{14}$$

Using Equation (14) in Equation (8) gives a third form for the generalized Ohm's law, which is

$$\vec{J} = \sigma \, \vec{E} - \beta \, \vec{J} \times \hat{k} \tag{15}$$

The cross product ($\vec{J} \times \hat{k}$) leads to a vector that has two non-trivial components along the $x$-axis and $y$-axis only, as

$$\vec{J} \times \hat{k} = \left(J_x \, \hat{i} + J_y \, \hat{j} + J_z \, \hat{k}\right) \times \hat{k} = -J_y \, \hat{i} + J_x \, \hat{j} \tag{16}$$

Using Equation (16) in Equation (15), and collecting terms for the three Cartesian components gives the following three scalar equations:

$$J_x = \sigma E_x - \beta J_y \tag{17}$$

$$J_y = \sigma E_y + \beta J_x \tag{18}$$

$$J_z = \sigma E_z \tag{19}$$

Given that the $z$-component of the within-channel electric field vector ($\vec{E}$) is zero here

$$E_z = 0 \tag{20}$$

then, Equation (19) implies also that the $z$-component of the electric current-density is zero

$$J_z = 0 \tag{21}$$

Equation (17) and Equation (18) are implicit expressions for the $x$-component and the $y$-component of the electric current-density, and these two components ($J_x$ and $J_y$) are clearly coupled. However, these two coupled equations can be solved simultaneously to obtain two explicit uncoupled expressions for these two components ($J_x$ and $J_y$), with the result being as follows:

$$J_x = \frac{\sigma}{1+\beta^2} (E_x - \beta E_y) \tag{22}$$

$$J_y = \frac{\sigma}{1+\beta^2} (E_y + \beta E_x) \tag{23}$$

The electric field vector within the plasma ($\vec{E}$) is the vector sum of the electric field source vector or the open-circuit electric field vector ($\vec{E}_{OC}$) that is induced due to the motion of the electrically-conductive plasma under the effect of the applied magnetic field vector, and the electric field sink ($\vec{E}_0$) that is consumed by the external electric load (if a load is connected). Thus,

$$\vec{E} = \vec{E}_0 + \vec{E}_{OC} \tag{24}$$

The load electric field ($\vec{E}_0$) in MHD power generation is dependent on the connected load. However, if the MHD channel is used in a reverse mode as a plasma accelerator (a magnetohydrodynamic drive), then this electric field ($\vec{E}_0$) can be then viewed as an applied one, and the electric current is applied to the MHD electrodes from a powerful external Direct Current (DC) source, rather than being collected from the MHD electrodes [146–148].

The induced or convection open-circuit electric field vector ($\vec{E}_{OC}$) is related to the bulk velocity of the plasma ($\vec{u}$) and the applied magnetic field vector ($\vec{B}$), while being perpendicular to both of them [149, 150]. Thus,

$$\vec{E}_{OC} = \vec{u} \times \vec{B} \qquad (25)$$

From the two previous equations, Equation (24) and Equation (25), we can write the following:

$$\vec{E} = \vec{E}_0 + \vec{u} \times \vec{B} \qquad (26)$$

For our case of a unidirectional magnetic field ($\vec{B} = B\,\hat{k}$) and unidirectional convective plasma velocity ($\vec{u} = u\,\hat{i}$), the *x*-component and the *y*-component of the above vector equation are

$$E_x = E_{0x} \qquad (27)$$

$$E_y = E_{0y} - u\,B \qquad (28)$$

Using the above two expressions for ($E_x$) and ($E_y$) in Equation (22) and Equation (23) gives

$$J_x = \frac{\sigma}{1+\beta^2}\,(E_{0x} - \beta\,E_{0y} + \beta\,u\,B) \qquad (29)$$

$$J_y = \frac{\sigma}{1+\beta^2}\,(E_{0y} - u\,B + \beta\,E_{0x}) \qquad (30)$$

These two scalar expressions are to be analyzed further, and their customized form is to be discussed for each case of the four designs of the linear Magnetohydrodynamic (MHD) channels. This is presented in the next four subsections, with one subsection dedicated to each channel design.

In the case of Faraday-type MHD linear channels (the continuous-electrode version or the segmented-electrode version), it is the *y*-component of the current density ($J_y$) and load electric field ($E_{oy}$) that are useful because these are the current densities effectively collected by the channel electrodes (the anode and cathode, which are separated vertically "along the *y*-axis"). In such cases, a Faraday load factor ($K_F$) can be introduced as the ratio of the load electric field ($E_{0y}$) to the induced electric field ($u\,B$). So, mathematically, we have

$$K_F \equiv \frac{E_{0y}}{u\,B} \qquad (31)$$

Equivalently, this Faraday load factor can be viewed as the ratio of the electric resistance of the load ($R_L$) to the total series resistance encountered by the electric current flow due to both the external electric resistance of the load and the effective internal electric resistance within the MHD generator itself, that is designated by the symbol ($R_G$) (this internal resistance is caused by the bulk plasma region, the boundary layer near the wall [151, 152], and any solid slag layer formed on the electrodes [153–156]). Therefore, the Faraday load factor ($K_F$) can be mathematically described as

$$K_F \cong \frac{R_L}{R_L + R_G} \qquad (32)$$

Therefore, the Faraday load electric field ($E_{0y}$) can be expressed as

$$E_{0y} = K_F \, u \, B \tag{33}$$

The value of ($K_F$) is bounded between 0 and 1. In the extreme condition of $K_F = 1$, the circuit is open (the external load is disconnected). This corresponds to setting the load resistance to infinity ($R_L = \infty$) in Equation (32). At the other extreme condition of $K_F = 0$, the circuit is shorted (the external load is replaced with a perfect electric conductor). This corresponds to setting the load resistance to zero ($R_L = 0$) in Equation (32). It can be shown that maximum power delivery to the "matched" external load occurs at an optimum Faraday load factor of $K_F = 0.5$ [157].

In general, the Direct Current (DC) electric power delivered to the load powered by a magnetohydrodynamic channel is [158]

$$P_w = -\iiint_{\text{volume}} \vec{J} \cdot \vec{E}_0 \, dV \tag{34}$$

where ($dV$) is an infinitesimal volume element.

The Direct Current (DC) electric power delivered to the load in the case of a Faraday-type MHD generator per unit volume of plasma is

$$P_F = E_{0y} \, | J_y | = K_F \, u \, B \, | J_y | \tag{35}$$

where $| J_y |$ is the absolute value of the vertical "i.e., parallel to the $y$-direction" current density (since it actually has a negative value due to pointing in the negative $y$-direction).

In the case of Hall-type linear MHD channels, it is the $x$-component of the current density ($J_x$) and load electric field ($E_{0x}$) that are useful because these are the current densities effectively collected by the channel electrodes (the anode and cathode, which are separated axially "along the $x$-axis"). In such cases, a Hall load factor ($K_H$) can be introduced as the ratio of the load electric field ($E_{0x}$) to the induced (open circuit) axial electric field ($\beta \, u \, B$). So, mathematically, we have

$$K_H \equiv \frac{|E_{0x}|}{\beta \, u \, B} \tag{36}$$

where $|E_{0x}|$ is the absolute value of the axial load electric field (since it is actually having a negative value due to pointing in the negative $x$-direction).

As in the case of the Faraday load factor, the Hall load factor can be viewed as the ratio of the electric resistance of the load ($R_L$) to the total series resistance encountered by the electric current flow due to both the external electric resistance of the load and the effective internal electric resistance within the MHD generator itself ($R_G$). Therefore, the Hall load factor ($K_H$) can be mathematically described as

$$K_H \cong \frac{R_L}{R_L + R_G} \tag{37}$$

Therefore, the absolute value of the Hall load electric field ($|E_{0x}|$) can be expressed as

$$|E_{0x}| = K_H \, \beta \, u \, B \qquad (38)$$

As in the case of the Faraday load factor, the value of ($K_H$) is bounded between 0 and 1. In the extreme condition of $K_H = 1$, the circuit is open. At the other extreme condition of $K_H = 0$, the circuit is shorted. Maximum power delivery to the "matched" external load occurs at an optimum Hall load factor of $K_H = 0.5$ [159].

The Direct Current (DC) electric power delivered to the load in the case of a linear Hall MHD generator per unit volume of plasma is

$$P_H = |E_{0x}| \, J_x = K_H \, \beta \, u \, B \, J_x \qquad (39)$$

## 3.2 Continuous-electrode Faraday channel

The first MHD channel design we review in this study is the continuous-electrode Faraday configuration. This is the simplest configuration among the four linear MHD channels in terms of physical construction and electric connectivity.

Using Equation (33) in Equation (29) and Equation (30) gives a customized form for the electric current-density components suitable for Faraday-type channels as

$$J_x = \frac{\sigma}{1+\beta^2} \, (E_{0x} - \beta \, K_F \, u \, B + \beta \, u \, B) = \frac{\sigma}{1+\beta^2} \, (E_{0x} + \beta \, u \, B \, [1 - K_F]) \qquad (40)$$

$$J_y = \frac{\sigma}{1+\beta^2} \, (K_F \, u \, B - u \, B + \beta \, E_{0x}) = \frac{\sigma}{1+\beta^2} \, (\beta \, E_{0x} - u \, B \, [1 - K_F]) \qquad (41)$$

In Figure 1, we illustrate the geometric layout of this channel in the $x - y$ plane. The top side represents the anode electrode, which has a negative (or grounded) polarity. The bottom side represents the cathode electrode, which has a positive polarity. In this sketch, the channel height (the separating distance between the two electrodes) is assumed to increase linearly with the axial distance ($x$). This is not necessary, and nonlinear profiles are permitted as well [160]. However, in the shown linearly-divergent channel, the equipotential surfaces (the virtual surfaces with constant electric potential $\Phi$) become flat planes (straight lines in the shown two-dimensional sketch). These potential planes are projected in the $x - y$ plane as inclined straight lines, with their inclination gradually change from being coincident with the anode at the top to being coincident with the cathode at the bottom. In our sketch, we provide arbitrary values for sample intermediate equipotential lines, in addition to the top anode (which is also an equipotential line), and the bottom cathode (which is an equipotential line). We assign an electric potential of 30 V to the cathode, and a reference zero potential to the anode. These are not realistic values because they are small (actual cathode potential can exceed hundreds of volts [161–163]), but they are provided just to improve the explanation through numerical examples.

The $y$-component of the load electric field ($E_{0y}$) is positive in the case of the continuous-electrode Faraday channel, meaning that it is pointing vertically up, from the bottom positive cathode to the top negative (or grounded reference) anode. The load electric field vector is in the direction of decreasing electric potential [164–166], and this justifies the upward direction of ($E_{0y}$).

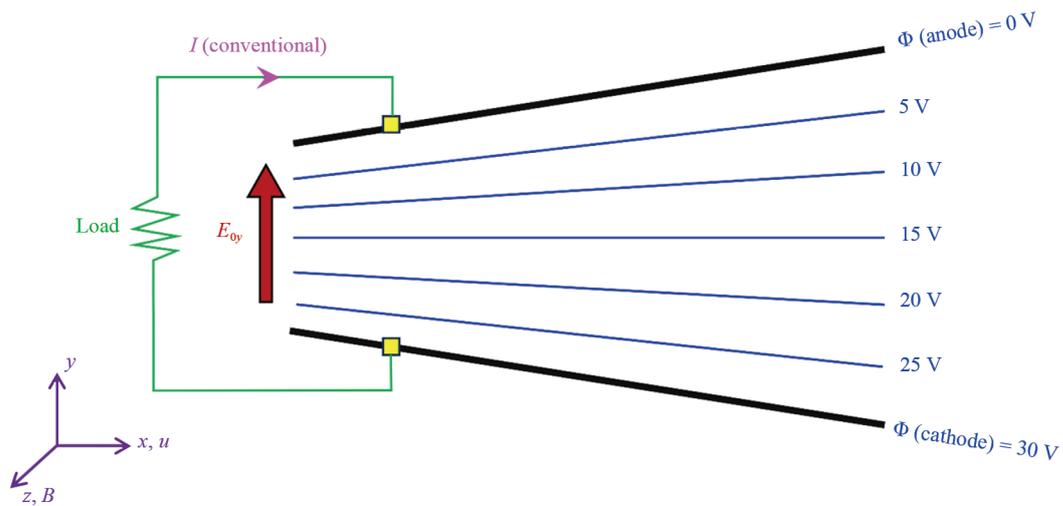

**Figure 1.** Graphical illustration of a linearly divergent continuous-electrode Faraday channel, with sample equipotential lines

In Figure 2, we further highlight the local direction of the load electric field vectors ($\vec{E}_0$), which is perpendicular to the local equipotential lines, making the load electric field vectors take the shape of circular arcs pointing from the bottom cathode to the top anode. Due to the anti-symmetry, the overall $x$-component of the load electric field vanishes, because in the upper half of the channel, the component ($E_{0x}$) is negative (upstream with respect to the moving plasma), while it is positive (downstream with respect to the moving plasma) in the lower half of the channel, as shown in the sketch.

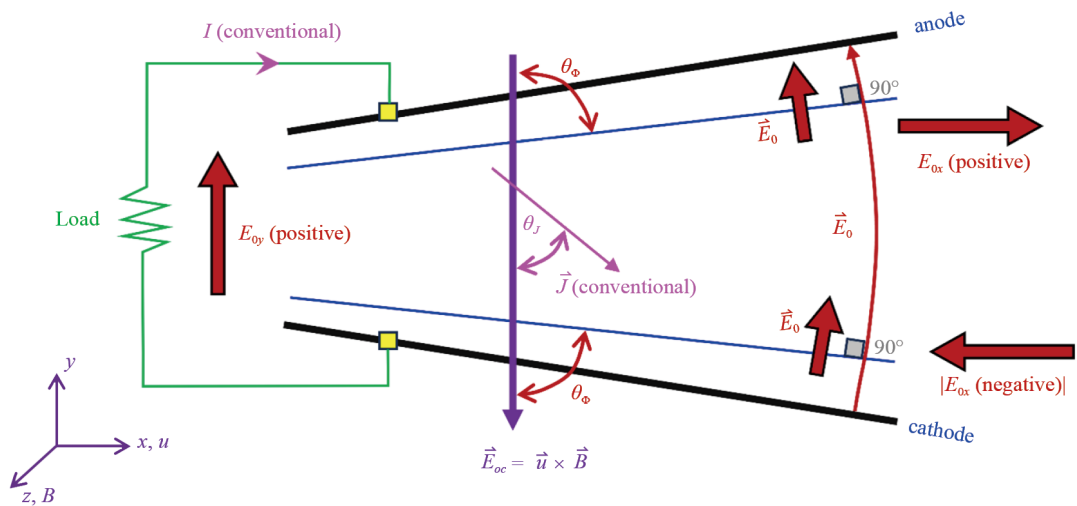

**Figure 2.** Graphical illustration of a continuous-electrode Faraday channel, with a demonstration of the direction of the current density and the load electric field

So, at the entire generator level, we have

$$E_{0x} = 0 \tag{42}$$

We also repeat below the general expression for ($E_{0y}$) for Faraday-type channels, which was presented earlier (Equation (33)).

If the MHD channel is uniform in height (no geometric divergence), then the condition ($E_{0x} = 0$) becomes applicable locally, not just at an integrated level.

Thus, the overall load electric field is effectively upward (from the bottom cathode to the top anode).

The absolute value of the local acute inclination angle ($\theta_\Phi$) of the equipotential line (measured from the "vertical" y-axis) is determined from the local values of the absolute x-component and the y-component (always positive here) of the load electric field; namely ($|E_{0x}|$) and ($E_{0y}$), respectively. Therefore, mathematically, we have

$$\theta_\Phi = \tan^{-1}\left(\frac{E_{0y}}{|E_{0x}|}\right) \tag{43}$$

For example, the centerline equipotential line is exactly horizontal, thus the local load electric field vector is exactly vertical (thus, $|E_{0x}| = 0$). Therefore, the centerline equipotential line has

$$\theta_\Phi = \tan^{-1}\left(\frac{E_{0y}}{|0|}\right) = \tan^{-1}(\infty) = 90° \tag{44}$$

Applying the condition ($E_{0x} = 0$) to the two base equations, Equation (40) and Equation (41), gives the following two customized relations for the electric current-density vectors in the case of the continuous-electrode Faraday channel:

$$J_x = \frac{\sigma}{1+\beta^2} \beta u B (1 - K_F) \tag{45}$$

$$J_y = -\frac{\sigma}{1+\beta^2} u B (1 - K_F) \tag{46}$$

Since the values of ($\sigma$), ($\beta$), ($u$), ($B$), and ($1 - K_F$) in the above equation are positive; the two previous equations imply that the x-component of the current density ($J_x$) is positive, while the y-component of the current density ($J_y$) is negative. Therefore, the electric current-density vector ($\vec{J}$) in the case of the continuous-electrode Faraday channel is inclined right down. Furthermore, the magnitude of this inclination angle ($\theta_J$), which is the acute angle measured from the vertical y-axis, is governed by

$$\theta_J = \tan^{-1}\left(\frac{J_x}{|J_y|}\right) \tag{47}$$

However, it can be seen from Equation (45) and Equation (46) that the ratio $\left(\frac{J_x}{|J_y|}\right)$ reduces to the Hall parameter ($\beta$). Therefore

$$\theta_J = \tan^{-1}(\beta) \tag{48}$$

Because the Hall parameter ($\beta$) depends on the electron mobility, which in turn depends on the local chemical composition, the temperature, and the pressure of the plasma gas [167–169]; the direction of the current density vectors $(\vec{J})$ may change spatially from one point to another within the MHD channel (but still pointing in the right-down direction). In the special case of uniform isothermal plasma, the value of $(\beta)$ becomes constant throughout the MHD channel, and thus the electric field vectors $(\vec{J})$ become parallel.

From Equation (35), the Direct Current (DC) electric power delivered to the load in the case of the continuous-electrode Faraday channel per unit volume of plasma $(P_{F-\text{cont}})$ is

$$P_{F-\text{cont}} : K_F \, u \, B \, |J_y| = K_F \, u \, B \, \frac{\sigma}{1+\beta^2} \, u \, B \, (1-K_F) = \frac{\sigma}{1+\beta^2} \, u^2 \, B^2 \, K_F \, (1-K_F) \tag{49}$$

When the above expression is optimized with respect to the Faraday load factor $(K_F)$, the optimum case occurs at $(K_F = 0.5)$. This means that the external resistance of the matched (optimized) load is equal to the internal resistance of the MHD generator, or

$$R_{L,\,\text{opt}-F} = R_G \tag{50}$$

In such a case of $(K_F = 0.5)$, the matched-load optimized power dissipation to the load (per unit plasma volume) in the continuous-electrode Faraday channel is

$$P_{F-\text{cont, opt}} = 0.25 \, \frac{1}{1+\beta^2} \, \sigma \, u^2 \, B^2 \tag{51}$$

The factor $\left(\frac{1}{(1+\beta^2)}\right)$ in the above equation represents a power penalty due to the uncollected "parasitic" Hall current density $(J_x)$.

In the limiting case of zero Hall effect $(\beta = 0)$, the above expression for the optimized load power (per unit plasma volume) becomes

$$P_{F-\text{cont, opt, ideal}} = 0.25 \, \sigma \, u^2 \, B^2 \tag{52}$$

However, practically, this ideal condition is not achievable with the continuous-electrode Faraday channel, because the theoretical condition that $(\beta = 0)$ also implies zero electron mobility $(\mu = 0)$ according to Equation (6). From Equation (7), such a condition implies zero electric conductivity $(\sigma = 0)$, and thus the MHD generator ceases to produce electricity.

The inevitable Hall effect loss associated with the continuous-electrode Faraday channel is a major disadvantage, making this channel type suitable only for a restricted regime of low $(\beta)$. For example, at a Hall parameter value of $(\beta = 0.5)$, 20% of the ideal power limit is lost, which is a reasonable loss; while at a Hall parameter of $(\beta = 1)$, 50% of the ideal power limit is lost; and at a Hall parameter of $(\beta = 2)$, 80% of the ideal power limit is lost and this is high. All these three values of $(\beta)$ are possible in open-cycle MHD generator plasma [170]. The use of continuous-electrode Faraday channel may be regarded as acceptable up to a limit of approximately $(\beta = 2)$ [171].

To avoid the aforementioned power loss problem, alternative MHD channel designs should be used, and this leads to the three alternative configurations of linear MHD channels to be discussed in the next three subsections.

We conclude this subsection by deriving an expression for the electric efficiency of the continuous-electrode Faraday channel. From the general expression in Equation (4), a reduced version can be obtained if the channel has uniform electromagnetic properties, and this reduced form is

$$\eta_{\text{elec}, F-\text{cont}} = \frac{|J_x\, E_{0x}| + |J_y\, E_{0y}|}{u\, |J_y|\, B} = \frac{|J_y|\, E_{0y}}{u\, |J_y|\, B} = \frac{E_{0y}}{u\, B} = \frac{K_F\, u\, B}{u\, B} = K_F \tag{53}$$

Thus, the electric efficiency reduces to the Faraday load factor ($K_F$) in the case of a continuous-electrode Faraday channel with uniform properties. This also means that the optimum electric efficiency ($\eta_{\text{elec}, F-\text{cont, opt}}$) in this channel configuration is 0.5 or 50%.

### 3.3 *Linear Hall channel*

After discussing the operation of the continuous-electrode Faraday MHD channel in the previous subsection, and highlighting the deficiency caused by the Hall effect in that type of linear MHD channels, rendering it to be undesirable at high Hall parameters exceeded unity; we discuss here the operation of an alternative type, which is the linear Hall MHD channel.

Unlike the continuous-electrode Faraday MHD channel, where a high Hall parameter ($\beta$) beyond unity renders that type undesirable; the linear Hall channel actually is designed for high values of the Hall parameter ($\beta$) (as high as 10 [172]), and it becomes undesirable at low values of ($\beta$).

Figure 3 illustrates the configuration of the linear Hall channel. Instead of separating the electrodes (anode and cathode) vertically, as was the case in the continuous-electrode Faraday channel; they are here separated axially. The positive cathode is at the rear of the channel, while the negative (or reference grounded) anode is at the front of the channel. Multiple vertical short-circuit links are inserted. Each vertical link is an equipotential line, and the electric potential increases downstream toward the cathode. We added some numerical example values, from 0 V at the anode to 30 V at the cathode (these are for explaining the variation of the electric potential, but they are not realistic values due to being very small) [173].

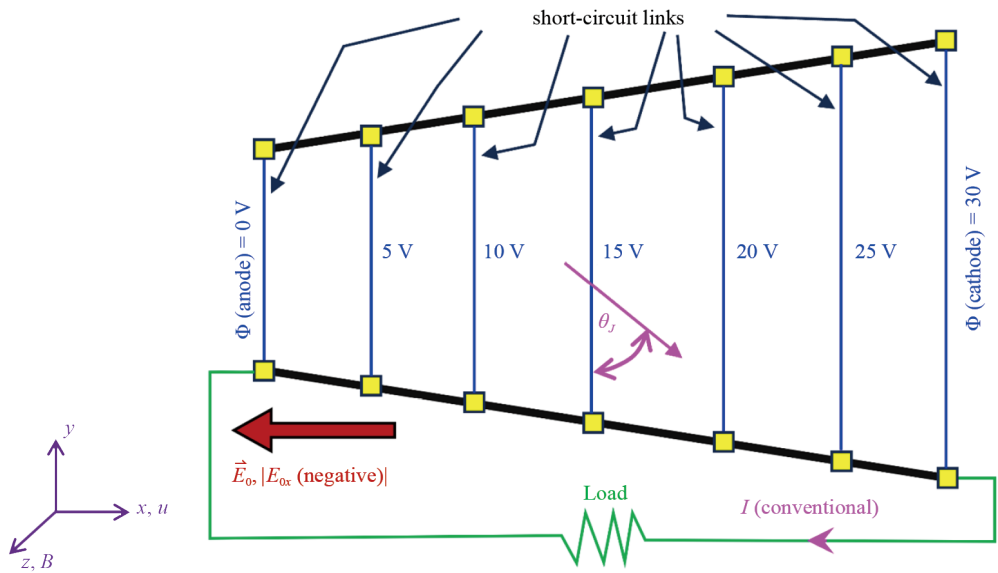

**Figure 3.** Graphical illustration of a linear Hall channel, with a demonstration of the direction of the current density and the load electric field

It can be seen in the sketch that the load electric field is purely horizontal, pointing upstream in the direction of decreasing electric potential ($\Phi$). The load electric field vector ($\vec{E}_0$) is the opposite of the gradient of electric potential ($\Phi$). Mathematically, this can be expressed as

$$\vec{E}_0 = -\nabla \Phi \tag{54}$$

Due to the vertical shorting introduced in the linear Hall MHD channel (making the equipotential lines vertical), the load electric field vector ($\vec{E}_0$) is purely horizontal, in the direction of the negative $x$-axis. Thus, the vertical component of the load electric field vector ($E_{0y}$) vanishes in the linear Hall channel. Therefore,

$$E_{0y} = 0 \tag{55}$$

$$\vec{E}_0 = E_{0x} \, \hat{i} = -|E_{0x}| \, \hat{i} \tag{56}$$

$$\theta_\Phi = \tan^{-1}\left(\frac{E_{0y}}{|E_{0x}|}\right) = \tan^{-1}\left(\frac{0}{|E_{0x}|}\right) = 0^\circ \tag{57}$$

Recognizing that ($E_{0x}$) has a negative value ($E_{0x} = -|E_{0x}|$), then from Equation (38), we can write

$$E_{0x} = -K_H \, \beta \, u \, B \tag{58}$$

Using the above equation and Equation (55) in Equation (29) and also in Equation (30), gives a customized form for the electric current-density components suitable for linear Hall channels as

$$J_x = \frac{\sigma}{1+\beta^2} \left(-K_H \, \beta \, u \, B + \beta \, u \, B\right) = \frac{\sigma}{1+\beta^2} \, \beta \, u \, B \, (1 - K_H) \tag{59}$$

$$J_y = \frac{\sigma}{1+\beta^2} \left(-u \, B - \beta^2 \, K_H \, u \, B\right) = -\frac{\sigma}{1+\beta^2} \, u \, B \, \left(1 + \beta^2 \, K_H\right) \tag{60}$$

The above two equations show that the current density vector in linear Hall channels has a positive $x$-component ($J_x = |J_x|$) but a negative $y$-component ($J_y = -|J_y|$). This is illustrated in the previous sketch, with the tilt angle (acute angle, measured from the $y$-axis) of the current-density vector ($\theta_J$) when measured from the vertical is mathematically expressed as

$$\theta_J = \tan^{-1}\left(\frac{J_x}{|J_y|}\right) = \tan^{-1}\left(\frac{\beta \, (1 - K_H)}{1 + \beta^2 \, K_H}\right) \tag{61}$$

In order for the current density vector ($\vec{J}$) to be parallel, the Hall parameter ($\beta$) has to be uniform, and this implies uniformity in the plasma thermo-chemical properties (chemical composition, temperature, and pressure).

From Equation (39), the Direct Current (DC) electric power delivered to the load per unit volume of plasma in the case of the linear Hall channel ($P_H$) is

$$P_H : K_H \, \beta \, u \, B \, J_x = K_H \, \beta \, u \, B \, \frac{\sigma}{1+\beta^2} \, \beta \, u \, B \, (1 - K_H) = \frac{\sigma}{1+\beta^2} \, u^2 \, B^2 \, \beta^2 K_H \, (1 - K_H) \tag{62}$$

Similar to the case of continuous-electrode Faraday channels, when the above expression is optimized with respect to the Hall load factor ($K_H$), the optimum case occurs at ($K_H = 0.5$). This means that the external resistance of the matched (optimized) load is equal to the internal resistance of the MHD generator, or

$$R_{L,\,\text{opt}-H} = R_G \tag{63}$$

In such a case of ($K_H = 0.5$), the matched-load optimized power dissipation to the load (per unit plasma volume) in the linear Hall channel is

$$P_{H,\,\text{opt}} = 0.25 \, \frac{\beta^2}{1+\beta^2} \, \sigma \, u^2 \, B^2 \tag{64}$$

Comparing this expression for ($P_{H,\,\text{opt}}$) to the one derived earlier for ($P_{F,\,\text{opt}}$) shows that the power penalty factor $\left(\frac{1}{(1+\beta^2)}\right)$ in the continuous-electrode Faraday becomes $\left(\frac{\beta^2}{(1+\beta^2)}\right)$ in the linear Hall channel. This Hall power penalty factor approaches unity (thus, the penalty diminishes) at high values of the Hall parameter ($\beta \gg 1$). This explains how the linear Hall channel is favored over the continuous-electrode Faraday for high ($\beta$). On the other hand, the continuous-electrode Faraday channel exhibits a smaller power penalty at ($\beta < 1$). The power penalty factors for both channel types become equal to 50% at ($\beta = 1$). The ratio of the matched-load volumetric power density for the linear Hall channel compared to the continuous-electrode Faraday channel is the square of the Hall parameter. Therefore,

$$\frac{P_{H,\,\text{opt}}}{P_{F,\,\text{opt}-\text{cont}}} = \beta^2 \tag{65}$$

To help quantify the advantage of continuous-electrode Faraday MHD generators at low ($\beta$), and the advantage of linear Hall MHD generators at high ($\beta$), we compare in Table 1 the power penalty factors for both types of MHD linear channels (continuous-electrode and Hall) for a wide range of ($\beta$) from 0 to 10. It is apparent that at ($\beta = 2$), the linear Hall channel is four times more useful than the continuous-electrode Faraday channel.

Table 1. Gain in the power output for the linear Hall channel compared to the continuous-electrode Faraday channel at different Hall parameters

| | Power penalty factor | | |
|---|---|---|---|
| Hall parameter | Continuous-electrode Faraday | Linear Hall | $\frac{P_{H,\,opt}}{P_{F,\,opt-cont}} = \beta^2$ |
| 0 | 100% | 0% | 0 |
| 0.25 | 94.1176% | 0.0588 | 0.0625 |
| 0.5 | 80% | 20% | 0.25 |
| 0.75 | 64% | 36% | 0.5625 |
| 1 | 50% | 50% | 1 |
| 1.25 | 39.0244% | 60.9756% | 1.5625 |
| 1.5 | 30.7692% | 69.2308% | 2.25 |
| 1.75 | 24.6154% | 75.3846% | 3.0625 |
| 2 | 20% | 80% | 4 |
| 2.5 | 13.793% | 86.2069% | 6.25 |
| 3 | 10% | 90% | 9 |
| 4 | 5.8824% | 94.1176% | 16 |
| 5 | 3.8462% | 96.1538% | 25 |
| 6 | 2.7027% | 97.2973% | 36 |
| 7 | 2% | 98% | 49 |
| 8 | 1.5385% | 98.4615% | 64 |
| 9 | 1.2195% | 98.7805% | 81 |
| 10 | 0.9901% | 99.0099% | 100 |

We point out here that in the linear Hall channel, because there are multiple electric connections (the shorting links) between the anode and cathode pairs, and each of these intermediate electric connections has an electric potential exceeding that of the anode; each of these intermediate links can be used as an intermediate cathode that powers a separate electric load (connected from the other terminal to the global anode at the entrance of the MHD channel). This possibility is illustrated in Figure 4. However, in the current study, we assume in the analysis the simple case of a single electric load connected between the MHD overall anode and overall cathode. This allows consistency when comparing this channel type with the continuous-electrode Faraday channel (which admits only a single electric load).

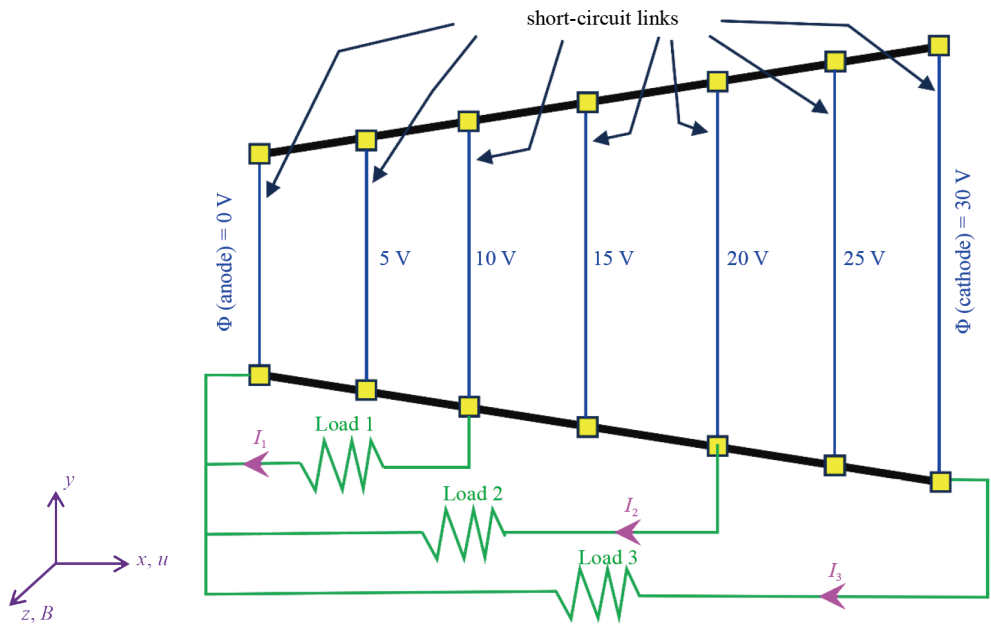

**Figure 4.** Graphical illustration of a linear Hall channel, in the case of three external loads being powered simultaneously

We conclude this subsection by deriving a special expression for the electric efficiency of the linear Hall channel. From the general expression in Equation (4), a reduced version can be obtained if the channel has uniform electromagnetic properties, and this reduced form is

$$\eta_{\text{elec}, H} = \frac{|J_x\, E_{0x}| + |J_y\, E_{0y}|}{u\, |J_y|\, B} = \frac{J_x\, |E_{0x}|}{u\, |J_y|\, B} = \frac{\dfrac{\sigma}{1+\beta^2}\, \beta\, u\, B\, (1-K_H)\, K_H\, \beta\, u\, B}{u\, \dfrac{\sigma}{1+\beta^2}\, u\, B\, (1+\beta^2\, K_H)\, B} \quad (66)$$

This can be simplified to

$$\eta_{\text{elec}, H} = \frac{\beta^2\, (1-K_H)\, K_H}{1+\beta^2\, K_H} = \frac{\beta^2}{1+\beta^2\, K_H}\, (1-K_H)\, K_H \quad (67)$$

At high Hall parameters ($\beta \to \infty$), the above expression approaches the following limit:

$$\eta_{\text{elec}, H}(\beta \to \infty) = 1 - K_H \quad (68)$$

At the optimum Hall load factor ($K_H = 0.5$), the electric efficiency expression in Equation (67) can be further simplified to

$$\eta_{\text{elec}, H}\,(K_H = 0.5) = 0.25\, \frac{\beta^2}{1+0.5\, \beta^2} = \frac{\beta^2}{4+2\, \beta^2} \quad (69)$$

which we donate as $(\eta_{\text{elec}, H, \text{opt}})$, which refers to the optimality condition of the Hall load factor.

At high Hall parameters ($\beta \to \infty$), the above expression approaches the following limit:

$$\eta_{\text{elec}, H}\left(K_H = 0.5, \ \beta \to \infty\right) = \frac{1}{2} \ \text{or} \ 50\% \tag{70}$$

## 3.4 *Segmented-electrode Faraday channel*

After discussing the operational conditions and performance of the continuous-electrode Faraday MHD channel and the linear Hall MHD channel, we here discuss a third configuration of MHD generator channels, which is the segmented-electrode Faraday channel.

In Figure 5, we provide a graphical illustration of the segmented-electrode Faraday channel, which clearly differs from both the continuous-electrode Faraday channel and the linear Hall channel. Like the continuous-electrode Faraday channel, the electrodes are separated vertically (along the *y*-axis); but unlike the continuous-electrode Faraday channel, each electrode (the bottom positive cathode and the top negative anode) is now divided into multiple electrode segments that are electrically insulated from the adjacent segments.

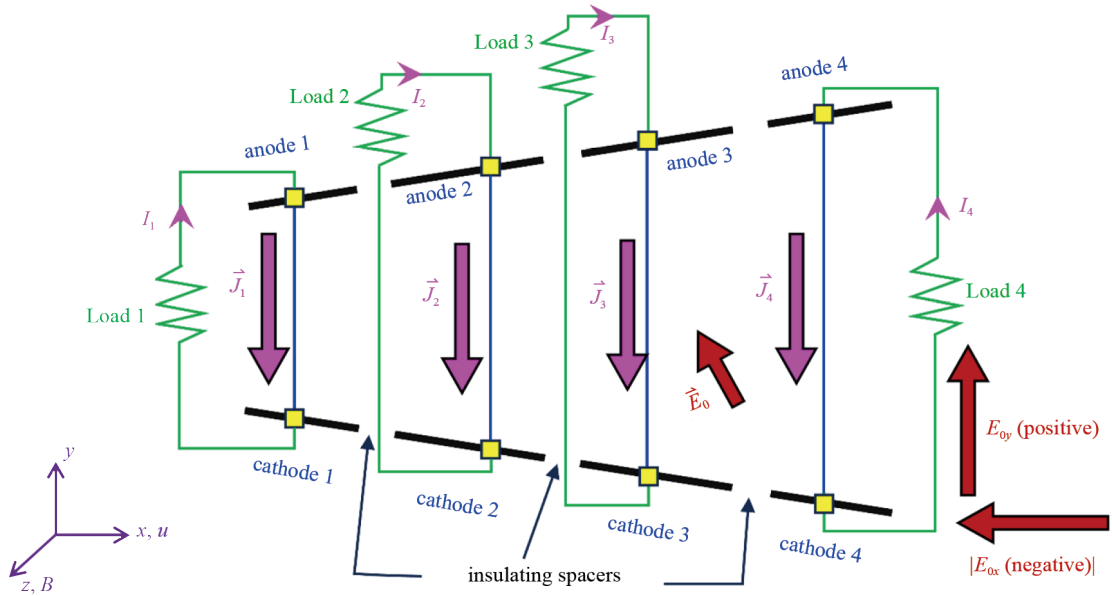

**Figure 5.** Graphical illustration of a segmented-electrode Faraday channel, with a demonstration of the direction of the current density and the load electric field

The motivation behind the segmented-electrode Faraday channel can be clarified by recalling the adverse effect of the Hall current density ($J_x$) in the continuous-electrode Faraday channel due to the Hall effect, which causes power loss for the continuous-electrode Faraday channel, in the form of a penalty factor $\left(\dfrac{1}{(1+\beta^2)}\right)$. Operating under a condition of vanishing Hall parameter ($\beta = 0$) eliminates is penalty because the penalty factor becomes unity. However, such a condition of zero Hall parameter also means zero electric conductivity, as implied by Equation (6) and Equation (7).

Therefore, an alternative method of avoiding the power loss under non-zero Hall parameter is to adapt the electric connectivity such that the Hall current density ($J_x$) vanishes, even with the presence of the unavoidable Hall effect.

In order to suppress the Hall current density, the top and bottom electrodes are segmented into multiple segments, and this arrangement does not give a chance for the Hall current density to develop. Ideally, there should be an infinite number of segments. However, a finite number of segments is practically possible. This resembles the lamination of a solid iron core in an electric transformer in order to suppress the undesirable but unavoidable eddy currents [174–176].

For each pair of opposite segments (cathode and anode), an external load is connected. This might be a drawback in this channel configuration, where having multiple individual loads may not represent the exact demand pattern.

When assuming that the Hall current density successfully vanishes, the following condition becomes a characteristic feature of the segmented-electrode Faraday channel:

$$J_x = 0 \tag{71}$$

Consequently, this means that the inclination angles of the electric current-density vectors ($\vec{J}$), measured from the vertical are zero. Therefore, each electric current-density vector ($\vec{J}$) is perfectly vertical (pointing down, from the top anode segment to the bottom cathode segment). Mathematically, this is expressed as

$$\theta_J = \tan^{-1}\left(\frac{J_x}{|J_y|}\right) = \tan^{-1}\left(\frac{0}{|J_y|}\right) = 0° \tag{72}$$

Therefore, in the segmented-electrode Faraday channel; the direction of the current density vectors is restricted to the vertical orientation. However, the load electric field vectors are not subject to such a constraint. In fact, the load electric field possesses an axial component ($E_{0x}$) and a vertical component ($E_{0y}$). Like the continuous-electrode Faraday channel, Equation (33) for ($E_{0y}$) is still valid.

This means that ($E_{0y}$) is positive (pointing upward toward the negative anodes).

However, the axial component of the load electric field vector ($E_{0x}$) is no longer zero in the segmented-electrode Faraday channel as was the case in the continuous-electrode Faraday channel. The expression for ($E_{0x}$) can be derived from its general expression in Equation (40). After setting ($J_x = 0$) in Equation (40), we obtain a condition on ($E_{0x}$) as

$$0 = \frac{\sigma}{1+\beta^2}\left(E_{0x} + \beta\, u\, B\, [1-K_F]\right) \tag{73}$$

This leads to the following expression for ($E_{0x}$) in the case of a segmented-electrode Faraday channel:

$$E_{0x} = -\beta\, u\, B\, (1-K_F) \tag{74}$$

This shows that ($E_{0x}$) is negative, which in turn means that the electric potential ($\Phi$) decreases in the axial direction, as the x-coordinate increases.

The absolute value of the local acute inclination angle ($\theta_\Phi$) of the equipotential lines (measured from the "vertical" y-axis) in the segmented-electrode Faraday channel can be described as

$$\theta_\Phi = \tan^{-1}\left(\frac{E_{oy}}{|E_{ox}|}\right) = \tan^{-1}\left(\frac{K_F u B}{\beta u B(1-K_F)}\right) = \tan^{-1}\left(\frac{K_F}{\beta(1-K_F)}\right) \tag{75}$$

Due to the dependence on the Hall parameter ($\beta$), the angle ($\theta_\Phi$) is not necessarily constant throughout the MHD channel, and thus the equipotential lines are not necessarily parallel. However, we illustrate them in Figure 6 in the special case where these equipotential lines are parallel straight lines (for simplicity), and we also show in this figure how the angle ($\theta_\Phi$) is defined.

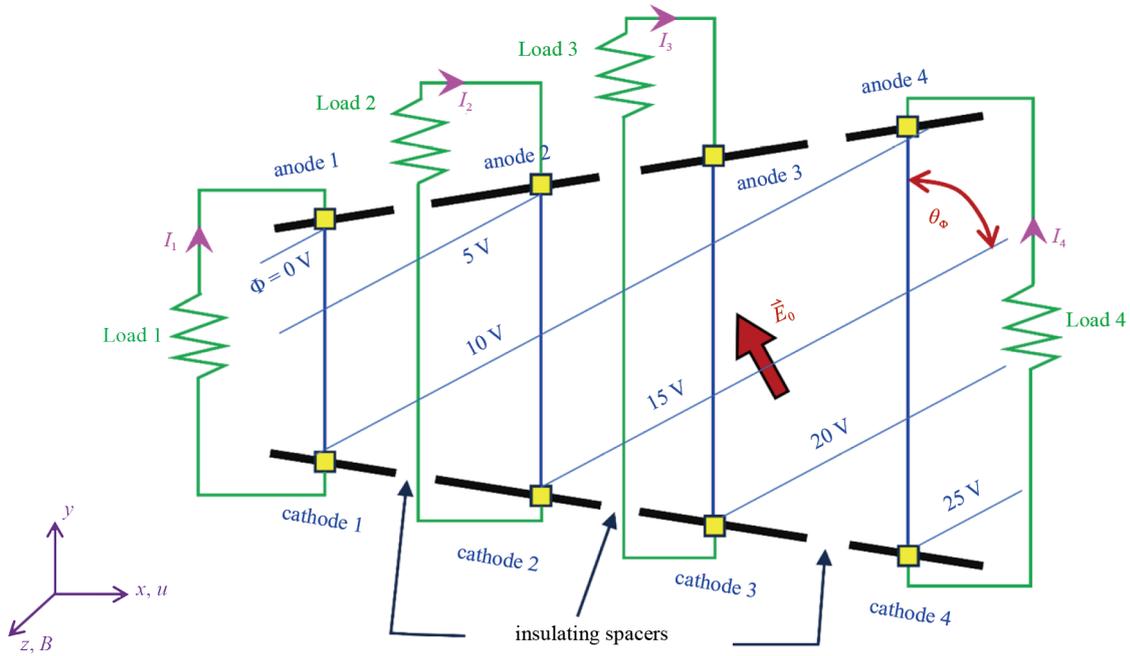

**Figure 6.** Graphical illustration of a segmented-electrode Faraday channel, with a demonstration of the direction of the current density and the load electric field

From Equation (41), the $y$-component of the current density in the segmented-electrode Faraday channel is obtained by using Equation (74), which gives:

$$J_y = \frac{\sigma}{1+\beta^2}\left(-\beta^2 uB[1-K_F] - uB[1-K_F]\right) = -\frac{\sigma}{1+\beta^2} uB(1-K_F)(1+\beta^2) \tag{76}$$

Thus,

$$J_y = -\sigma uB(1-K_F) \tag{77}$$

This shows that the electric current-density is downward (having a negative sign).

From Equation (35), the Direct Current (DC) electric power delivered to the load (treating the multiple connected loads as a single one) in the case of the segmented-electrode Faraday channel per unit volume of plasma ($P_{F-\text{seg}}$) is:

$$P_{F-\text{seg}} = K_F uB|J_y| = K_F uB\sigma uB(1-K_F) = \sigma u^2 B^2 K_F(1-K_F) \tag{78}$$

By comparing this expression for ($P_{F-\text{seg}}$) to the counterpart expression of ($P_{F-\text{cont}}$) in the case of the continuous-electrode Faraday channel in Equation (49), it is evident that the penalty factor of $\left(\dfrac{1}{(1+\beta^2)}\right)$ does not appear in the case of the segmented-electrode Faraday channel. Thus, the electrode segmentation in the Faraday-type channel is successful in making the channel performance independent of the Hall effect, but multiple electrode pairs and loads replace the simpler configuration of a single electrode pair and single load in the case of the continuous-electrode Faraday channel.

As in the case of the continuous-electrode Faraday channel, the optimized output Direct Current (DC) power to the load occurs at a matched load with ($K_F = 0.5$) or ($R_{L,\,\text{opt}-F} = R_G$). The optimized (matched-load) volumetric power density in the case of the segmented-electrode Faraday channel is:

$$P_{F-\text{seg, opt}} = 0.25 \sigma u^2 B^2 \tag{79}$$

This is the same as the ideal (at the theoretical limit of vanishing Hall parameter) volumetric power density for the continuous-electrode Faraday channel with optimized (matched) load in Equation (52).

Comparing this expression for ($P_{F-\text{seg, opt}}$) to the expression of ($P_{H,\,\text{opt}}$) in the case of the linear Hall channel in Equation (64), it becomes clear that the penalty factor of $\left(\dfrac{\beta^2}{(1+\beta^2)}\right)$ no longer appears for the case of the segmented-electrode Faraday channel.

We conclude this subsection by deriving a special expression for the electric efficiency of the segmented-electrode Faraday channel. From the general expression in Equation (4), a reduced version can be obtained if the channel has uniform electromagnetic properties, and this reduced form is:

$$\eta_{\text{elec},\,F-\text{seg}} = \frac{|J_x E_{0x}| + |J_y E_{0y}|}{u|J_y|B} = \frac{|J_y|E_{0y}}{u|J_y|B} = \frac{E_{0y}}{uB} = \frac{K_F uB}{uB} = K_F \tag{80}$$

Thus, the electric efficiency reduces to the Faraday load factor ($K_F$) in the case of a segmented-electrode Faraday channel with uniform properties. This is the same result obtained for the continuous-electrode Faraday channel. This also means that the optimum electric efficiency ($\eta_{\text{elec},\,F-\text{seg, opt}}$) in this channel configuration is 0.5 or 50%.

## 3.5 *Diagonal-electrode channel*

In the previous subsection, we showed how the segmented-electrode Faraday Magnetohydrodynamic (MHD) channel possesses desirable performance through ideal utilization of the MHD volume without being affected by the Hall parameter. However, we showed that this ideal condition comes at the expense of complicating the construction and electric connectivity, while a large number (theoretically infinite number) of anode-cathode pairs are needed for powering a large number of individual loads.

It is desirable to design a fourth configuration of linear MHD channels that maintains the excellent performance of the segmented-electrode Faraday channel while being less demanding in terms of the construction complexity and being able to power a single load if wanted. We here discuss this design, referred to as the (diagonal-electrode channel) or the (diagonal channel).

In the diagonal-electrode MHD channel, the suppression of the Hall current is achieved, but not through complicated segmentation as in the case of the segmented-electrode Faraday channel. Rather, this condition of ($J_x = 0$) attained through manipulating the direction of the electric field vectors such that their direction is the same as those implied in the segmented-electrode Faraday channel. This also means that the inclination angle of the equipotential lines (measured from the vertical) should also match the one found in the segmented-electrode Faraday channel.

It is useful to repeat Equations (75) here the mathematical expression arrived at in the previous subsection for the absolute value of the local acute inclination angle ($\theta_\Phi$) of the equipotential line (measured from the "vertical" *y*-axis) in the segmented-electrode Faraday channel.

This means that the tangent of the equipotential inclination is:

$$\tan(\theta_\Phi) = \frac{K_F}{\beta(1 - K_F)} \tag{81}$$

The above expression can be manipulated to derive a mathematical expression for the Faraday load factor ($K_F$) as a function of the equipotential lines angle ($\theta_\Phi$). The result is:

$$K_F = \frac{\beta \tan(\theta_\Phi)}{1 + \beta \tan(\theta_\Phi)} \tag{82}$$

The control of the direction of the equipotential lines in the diagonal channel is achieved by introducing inclined short-circuit links, tilted at the desired inclination angle ($\theta_\Phi$) from the vertical as shown in Figure 7. In this sketch, we assume a variation of the electric potential from 0 V at the MHD anode (located at the entrance of the MHD channel) to 20 V at the MHD cathode (located at the rear of the MHD channel). We also assume a single external load connected between these two primary electrodes (the primary anode and the primary cathode).

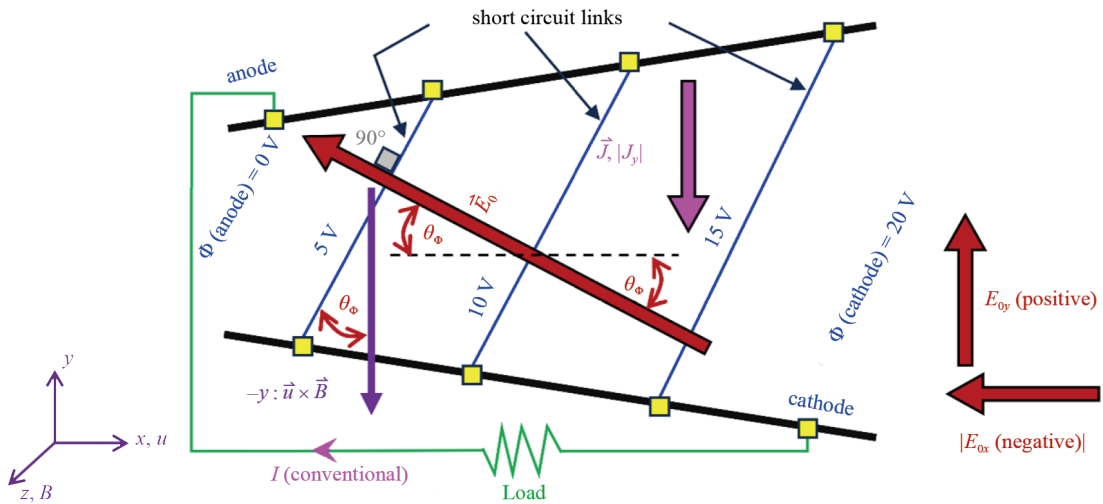

**Figure 7.** Graphical illustration of a diagonal-electrode channel, with a demonstration of the direction of the current density and the load electric field

From Equation (75), we can also extract a geometric condition of the load electric field vector ($\vec{E}_0$), whose negative axial component ($E_{0x}$) and its positive vertical component ($E_{0y}$) should be related according through the angle ($\theta_\Phi$) as:

$$E_{0x} = -\frac{E_{0y}}{\tan(\theta_\Phi)} = -E_{0y} \cot(\theta_\Phi) \tag{83}$$

As in the Faraday-type channels, it's given as Equation (33).
Therefore, in the diagonal MHD channel, the (negative) axial component of the load electric field can be expressed using Equation (83) and Equation (33) as:

$$E_{0x} = -\frac{K_F uB}{\tan(\theta_\Phi)} = -K_F uB \cot(\theta_\Phi) \tag{84}$$

Using the expression of ($E_{0x}$) in Equation (84) and the expression of ($\tan(\theta_\Phi)$) in Equation (81) gives another form for expressing the axial component ($E_{0x}$), which is:

$$E_{0x} = -\frac{K_F uB}{\frac{K_F}{\beta(1-K_F)}} = -uB\beta(1-K_F) \tag{85}$$

It is useful now to repeat the general expression of the axial component ($J_x$) of the electric current-density vector ($\vec{J}$) for Faraday-type channels (Equation (40)).

Using the expression for the target profile of ($E_{0x}$) for a diagonal channel, as provided through Equation (85), in the above equation for ($J_x$) shows that the component ($J_x$) automatically vanishes, as follows:

$$J_x = \frac{\sigma}{1+\beta^2}(-uB\beta[1-K_F] + uB\beta[1-K_F]) = \frac{\sigma}{1+\beta^2} uB\beta(0) = 0 \tag{86}$$

The vanishing of the axial component of the electric current-density vectors ($\vec{J}$) means that these vectors are exactly vertical (parallel to the y-axis). Consequently, the inclination angles of these vectors ($\theta_J$), measured from the vertical, are zero. This situation is identical to the one obtained in the segmented-electrode Faraday channel (Equation (72)).

We also repeat the general expression of the vertical component ($J_y$) of the electric current-density vector ($\vec{J}$) for Faraday-type channels (Equation (41)).

Using the earlier expression for ($E_{0x}$), as given in Equation (85), shows that for the diagonal channel, the component ($J_y$) becomes:

$$J_y = \frac{\sigma}{1+\beta^2}(-uB\beta^2[1-K_F] - uB[1-K_F]) = -\frac{\sigma}{1+\beta^2} uB[1-K_F](1+\beta^2) \tag{87}$$

Thus, for the diagonal channel, we have the same expression for ($J_y$) as the one reached earlier for the segmented-electrode Faraday channel (Equation (77)).

The Direct Current (DC) electric power delivered to the load (treating the multiple connected loads as a single one) in the case of the diagonal-electrode channel per unit volume of plasma ($P_D$) is the same as the one provided earlier for the segmented-electrode Faraday channel ($P_{F-\text{seg}}$). The mathematical expression for ($P_D$) is:

$$P_D = K_F uB|J_y| = K_F uB\sigma uB(1-K_F) = \sigma u^2 B^2 K_F(1-K_F) \tag{88}$$

Therefore, neither the penalty factor $\left(\frac{1}{(1+\beta^2)}\right)$ nor the penalty factor $\left(\frac{\beta^2}{(1+\beta^2)}\right)$ appear in the present case of a diagonal-electrode channel.

As in the case of the continuous-electrode Faraday channel and the segmented-electrode Faraday channel, the optimized value of ($K_F$) is 0.5; and optimized power density in the case of the diagonal-electrode channel is:

$$P_{D,\,\text{opt}} = 0.25\sigma u^2 B^2 \tag{89}$$

We would like to add three remarks about the diagonal MHD channel.

The first remark is that, despite the attractive performance of the diagonal channel as described above, it should be noted that this is constrained to a particular value of ($K_F$) and a corresponding uniform value of ($\beta$). In reality, it is difficult to maintain such a specific operating point (on-design operation), and thus operating at off-design regimes is likely to happen [177–179]. In the off-design condition, the expressions we provided for the diagonal channel break, as these assume a perfect design point.

The second remark is that the diagonal-electrode channel reduces to a linear Hall channel in the special value of ($\theta_\Phi = 0°$), which corresponds to vertical shorting links.

The third remark is that, as was the case for the linear Hall channel, the diagonal-electrode channel permits powering multiple loads simultaneously because there are multiple electric connections (the shorting links) that offer multiple levels of electric potential. This possibility is illustrated in Figure 8.

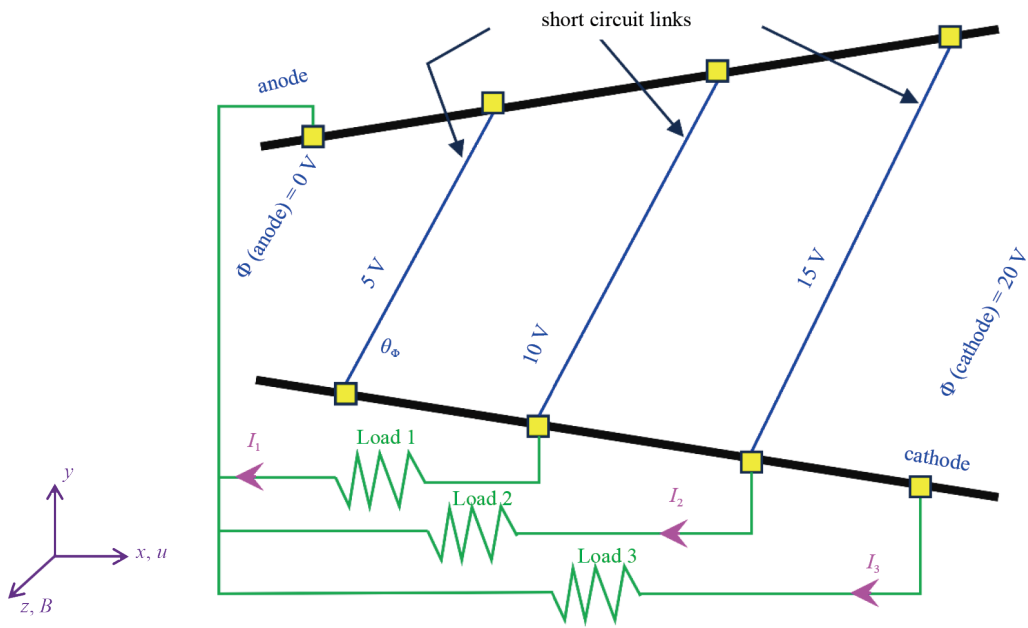

**Figure 8.** Graphical illustration of a diagonal-electrode channel, in the case of three external loads being powered simultaneously

As done for the previous three subsections, we conclude this subsection by deriving a special expression for the electric efficiency of the diagonal-electrode channel. From the general expression in Equation (4), a reduced version can be obtained if the channel has uniform electromagnetic properties and it is operating at its design point, and this reduced form is:

$$\eta_{\text{elec},D} = \frac{|J_x E_{0x}| + |J_y E_{0y}|}{u|J_y|B} = \frac{|J_y|E_{0y}}{u|J_y|B} = \frac{E_{0y}}{uB} = \frac{K_F uB}{uB} = K_F \tag{90}$$

Thus, the electric efficiency is reduced to ($K_F$). This is the same result obtained for either the continuous-electrode Faraday channel or the segmented-electrode Faraday channel. This also means that the optimum electric efficiency ($\eta_{\text{elec}, D, \text{opt}}$) in this channel configuration is 0.5 or 50%.

## 4. Discussion

The purpose of the current section is to augment the presented results by adding two supplementary topics. The first topic is a summary of key equations describing various performance variables for each linear MHD channel type, as derived in detail in the previous section. The second topic is a numerical set of values for some of these key parameters, which is useful for bridging the gap between the preceding theoretical analysis and real applications.

### 4.1 *Summary of MHD channel performance expressions*

In this subsection, we compare in Table 2 various key characteristics of the continuous-electrode Faraday channel and the linear Hall channel, whose geometric and electric connectivity are very different.

**Table 2.** Comparison between two types of linear magnetohydrodynamic channels

| Quantity | Continuous-electrode Faraday | Linear Hall |
| --- | --- | --- |
| $E_{0x}$ | 0 | $-K_H \beta u B$ |
| $E_{0y}$ | $K_F u B$ | 0 |
| $\theta_\Phi$ | 90° | 0° |
| $J_x$ | $\frac{\sigma}{1+\beta^2} \beta u B (1-K_F)$ | $\frac{\sigma}{1+\beta^2} \beta u B (1-K_H)$ |
| $J_y$ | $-\frac{\sigma}{1+\beta^2} u B (1-K_F)$ | $-\frac{\sigma}{1+\beta^2} u B (1+\beta^2 K_H)$ |
| $\theta_J$ | $\tan^{-1}(\beta)$ | $\tan^{-1}\left(\frac{\beta(1-K_H)}{1+\beta^2 K_H}\right)$ |
| $P$ | $\frac{\sigma}{1+\beta^2} u^2 B^2 K_F (1-K_F)$ | $\frac{\sigma}{1+\beta^2} u^2 B^2 \beta^2 K_H (1-K_H)$ |
| $P_{\text{opt}}$ | $0.25 \frac{1}{1+\beta^2} \sigma u^2 B^2$ | $0.25 \frac{\beta^2}{1+\beta^2} \sigma u^2 B^2$ |
| $\eta_{\text{elec}}$ | $K_F$ | $\frac{\beta^2 (1-K_H) K_H}{1+\beta^2 K_H}$ |
| Number of loads | 1 | 1 or more |

Then in Table 3, we provide a similar comparison, but between the segmented-electrode Faraday channel and its performance-equivalent diagonal-electrode channel (when operating at is design point). It can be seen that these two types are very similar in terms of their operation.

Table 3. Comparison between the other two types of linear magnetohydrodynamic channels

| Quantity | Segmented-electrode Faraday | Diagonal-electrode |
|---|---|---|
| $E_{0x}$ | $-\beta uB(1-K_F)$ | Same as segmented-electrode Faraday |
| $E_{0y}$ | $K_F uB$ | Same as segmented-electrode Faraday |
| $\theta_\Phi$ | $\tan^{-1}\left(\dfrac{K_F}{\beta(1-K_F)}\right)$ | Same as segmented-electrode Faraday |
| $J_x$ | 0 | Same as segmented-electrode Faraday |
| $J_y$ | $-\sigma uB(1-K_F)$ | Same as segmented-electrode Faraday |
| $\theta_J$ | $0°$ | Same as segmented-electrode Faraday |
| $P$ | $\sigma u^2 B^2 K_F(1-K_F)$ | Same as segmented-electrode Faraday |
| $P_{\text{opt}}$ | $0.25\sigma u^2 B^2$ | Same as segmented-electrode Faraday |
| $\eta_{\text{elec}}$ | $K_F$ | Same as segmented-electrode Faraday |
| Number of loads | multiple | 1 or more |

## 4.2 Numerical examples of linear MHD channel performance quantities

In the current subsection, we provide a quantitative estimate of some performance quantities presented above, based on typical values that can be encountered in real applications. This is summarized in Table 4.

Table 4. Magnitudes of important physical quantities in linear MHD channels at selected representative conditions

| Quantity | Magnitude | Reference |
|---|---|---|
| $\sigma$ | 5 S/m | [180–185] |
| $u$ | 2,000 m/s | [186–188] |
| $B$ | 5 T | [189–194] |
| $\beta$(F-cont) | 1 | [195, 196] |
| $K_F$(F-cont) | 0.5 | The current study itself |
| $E_{0y}$(F-cont) | 5,000 V/m (5 kV/m) | The current study itself |
| $|J_y$(F-cont)$|$ | 12,500 A/m$^2$ (12.5 kA/m$^2$) | The current study itself |
| $P_{F-\text{cont, opt}}$ | 62,500,000 W/m$^3$ (62.5 MW/m$^3$) | The current study itself |
| $\eta_{\text{elec}, F-\text{cont, opt}}$ | 50% | The current study itself |
| $\beta$(Hall) | 5 | [197–200] |
| $K_H$(Hall) | 0.5 | The current study itself |
| $|E_{0x}$(Hall)$|$ | 25,000 V/m (25 kV/m) | The current study itself |
| $J_x$(Hall) | 4,807.6923 A/m$^2$ (4.808 kA/m$^2$) | The current study itself |
| $P_{H, \text{opt}}$ | 120,192,300 W/m$^3$ (120.19 MW/m$^3$) | The current study itself |
| $\eta_{\text{elec}, H, \text{opt}}$ | 46.30% | The current study itself |
| $K_F$(F-seg) | 0.5 | The current study itself |
| $E_{0y}$(F-seg) | 5,000 V/m (5 kV/m) | The current study itself |
| $|J_y$(F-seg)$|$ | 25,000 A/m$^2$ (25 kA/m$^2$) | The current study itself |
| $P_{F-\text{seg, opt}}$ | 125,000,000 W/m$^3$ (125 MW/m$^3$) | The current study itself |
| $\eta_{\text{elec}, F-\text{seg, opt}}$ | 50% | The current study itself |
| $P_{D, \text{opt}}$ | 125,000,000 W/m$^3$ (125 MW/m$^3$) | The current study itself |
| $\eta_{\text{elec}, D, \text{opt}}$ | 50% | The current study itself |

In relation to recent studies, some of the above numerical estimations appear compatible. For example; atmospheric air-fuel combustion plasma with 0.125% cesium seeding (mole fraction) was found to enable an optimized ideal volumetric power density of 129.51 MW/m$^3$ for benchmarking [201], which is close to the above estimation of 125 MW/m$^3$. In another study; atmospheric air-fuel combustion plasma was shown to allow an optimized ideal volumetric power density of 104.49 MW/m$^3$ with 1% potassium seeding by mole [202], which is also close to the above estimation of 125 MW/m$^3$.

It is useful to add here that our analysis suggests that the volumetric power density depends linearly on the plasma electric conductivity. However, it depends quadratically on the magnetic-field flux density, the plasma speed, or the load factor.

## 5. Conclusions

In the current study, we provided a detailed mathematical analysis of the four main types of linear Magnetohydrodynamic (MHD) channels for power generation applications. Namely, these are the (1) continuous-electrode Faraday channel, (2) segmented-electrode Faraday channel, (3) linear Hall channel, and (4) diagonal-electrode channel. Through applying some assumptions (unidirectional applied magnetic field, unidirectional plasma velocity, low magnetic Reynolds number, and two-dimensional electric field), closed-form analytical expressions were derived to describe the operation and power generation performance of these four channel types. The study is enriched by contrasting the sets of equations governing various performance variables for each of the four linear MHD channel types, as well as by providing quantitative estimates to demonstrate the possible magnitude of key performance quantities of these four linear MHD channel types.

## Data availability statement

The data that support the findings of this study are available within the article itself.

## Conflict of interest

The author declares that he has no known competing financial interests or personal relationships that could have appeared to influence the work reported in this paper.